# A Conceptual Framework for Successful E-commerce Smartphone Applications: The Context of GCC

[1]**Adel Bahaddad,** [2]**Rayed AlGhamdi,** [2]**Seyed M. Buhari,** [2]**Madini O. Alassafi and** [2]**Ahmed Alzahrani**

[1]Information System Department, Faculty of Computing and Information Technology,
King Abdulaziz University, Jeddah, Saudi Arabia
E-mail: dbabahaddad10@kau.edu.sa

[2]Information Technology Department, Faculty of Computing and Information Technology
King Abdulaziz University, Jeddah, Saudi Arabia
E-mail: {raalghamdi8, mesbukary, malasafi, aalzahrani8}@kau.edu.sa

## ABSTRACT

Rapid expansion of online business has engulfed the GCC region. Such an expansion causes competition among business entities, causing the need to identify the factors that the customers use to choose a suitable mobile business application. Instead of just focusing on the visitors/users of the application, a shift in focus towards transforming casual customers to loyal customers is needed. The IS Success Model, whose main constructs are Information Quality, Quality Systems, Service Quality, User Satisfaction, Intention to Use and Net Benefits, includes diversified indicators along with their measures. This research considers User Satisfaction, Intention to Use and Net Benefits constructs as it is, but modified System Quality, Information Quality and Service Quality constructs based on previous state-of-the-art literature. The developed theoretical model was further tested surveying 803 GCC participants. Responses were analyzed using exploratory and confirmatory factor analysis. Study reveals the significance of Service Quality (consisting of M-loyalty Building, Customer Chat and feedback, Help and technical support, and Credibility and Reliability Build) over Information Quality and System Quality, impacting the importance of User Satisfaction over the other constructs.

## 1. INTRODUCTION

Many commercial enterprises, globally, have integrated e-commerce models in their businesses to obtain advantages of competing online [1]. Since 2000, e-commerce has undergone speedy growth in developed countries. Global e-commerce amounted to USD 0.27 trillion in 2000 and has reached over USD 2.356 trillion currently [2]. The United States (US) and Europe earn the major portion of global e-commerce revenue, while the Middle East and African region earnings are still relatively low [3]. Few years ago, when we studied the e-commerce adoption by retailers in KSA, few numbers of the participated sample were involved in activities of e-commerce [4]. Today, we witness that more than half of that sample sell their products online. The involvement in e-commerce for commercial enterprises is no longer an option. Either you sell online or lose your market share, powered by smartphone access. Extensive literature has studied the slowness of the e-commerce adoption in the Arab region [5-13]. The focus is on how to integrate and implement successful online business. Implementing what is successful in other countries is not guaranteed to work in a different country or culture. A popular example of that is the failure of eBay in China [14]. You cannot simply 'copy' eBay because it is successful in USA and 'paste' it elsewhere and expect it to work successfully. It requires to consider the culture differences. For this reason, this study take place. It focuses on GCC because it has notable growth of e-commerce activities [15]. Since so many businesses have started to involve in e-commerce activities and sell online, this study takes place to guide these businesses to consider the requirements to adopt the smartphone e-commerce applications successfully.

## 2. LITERATURE REVIEW AND THE DEVELOPED MODEL

During the last two decades, major companies and organisations have invested extensively in the implementation, evaluation and success of e-commerce systems to obtain a competitive advantage. Some questions may be raised to know the factors that affect consumers' intention to consistently choose a commercial site and application over another, and what quality level is acceptable for the consumer to continue to use the commercial application. Consumers generally want to use a specific site or application that contains most of their requirements for online purchasing before making a purchase decision [16]. Nowadays, e-commerce consumers have various choices and can switch between different merchants easily with a finger click. Therefore, providing fundamental requirements for consumers to purchase online may not be enough to ensure success. Moreover, online purchasing success does not depend on the consumers visiting their application, purchasing products or services, or registering as members with the application, but it is vital for the companies is to transform their casual consumers into loyal consumers. Thus, an e-commerce application must be integrated in terms of technical and human behavioral requirements in an appropriate manner [16]. As a consequence, the increase in the use of e-commerce technological advances has led to significant changes in the e-commerce field [17]. Nowadays, e-commerce and its success remain an important issue for anyone interested in developing electronic systems, from those on the Board of Directors of major companies to those working in research positions.

One of the popular models that studies a system's success is the Information Systems (IS) success model. The IS success model assumes that a system success is determined by six critical dimensions. These dimensions are system, information, and service qualities, intent to use, user satisfaction and net benefits, see Figure 1 [18]. System Quality (SQ) focuses on the associated requirements with application system, such as flexibility, reliability and ease of use. Information Quality (IQ) emphasizes on the information accuracy that is required to accomplish the online purchasing transaction through M-commerce applications, such as products description and information provided. Service Quality (SQ) discusses technical support features and service provided to customers, such as response time, reliability, accuracy and technical competence. Intent to Use (IU) measures the intention to use the actual system and willingness to engage in the future. User satisfaction (US) focuses on the level of satisfying users that use the system. Finally, the last dimension of the IS success model is Net Benefits (NB) that associates with benefits and values that reflect positively on the individual, community, or an organization. These are offerings, reduce costs, and enhance market efficiency, and improve the productivity for decision-making [18-20].

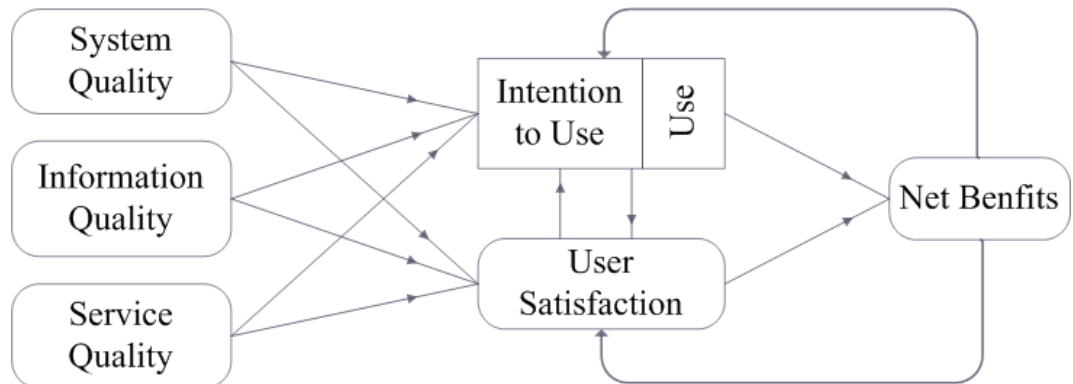

Figure 1: IS Success Model (DeLone & McLean, 2003)

The IS success model is a multi-dimensional model of its inputs, which in turn provides a multi-dimensional way of measurement. Therefore, the quality aspects of the model are not considered as being fixed-scale aspects from one model to another, and each research field has its own way of measuring of IS Success Model dimensions. These differences are advantageous, because each IS success research project has its own status of variable measurements [18, 19, 21]. As an example, variations in the methods of measurement have been used for the Information Quality dimension. Some studies have focused on many indicators such as content quality, delivery mechanism and quality of the information presented as important elements in the measurement of this construct [22]. Others slightly modified its focus from content quality to other elements, such as methods of attracting customers and ensuring the information presented is understandable [23]. Furthermore, certain system quality measurements, such as good user interface, flexibility in processing and change patterns, ability to adapt to new requirements, and ease of use, are critical aspects of systems quality measurement. The main goal in this research is not about exploring the complex dimensions behind designing systems quality requirements in commercial applications but exploring and studying what procedures related to building constructs of commercial applications should be adopted [23]. Finally, other research has defined the Service Quality measurements with a focus on parameters such as building loyalty and credibility, identifying ways to communicate with customers and solving problems that customers face before, during and after online transactions [24-26].

In our current study, we will adopt using the IS success model. However, we need to determine how the system, information and service qualities can be measured in designing e-commerce smartphones applications. The literature identified what can be called 'indicators' that distinguish good/bad e-commerce applications. These indicators can be divided into six groups: appearance, organization, content, customer-focus, interaction, and assurance. The appearance of applications (format, beauty, color usages, images fitting, font size and style, etc) plays significant role in attracting consumers to engage with the applications [27-29]. Designing an eye friendly application is vital only with valuable content. Providing updated information regularly, highlighting new items, viewing prices, offers and discounts, presenting purchase, shipping, return and after sales support policies, and the adaptation of using of local languages are all critical in determining content value [27, 30-34]. The provided content should be organized. Well-organized content positively correlates with the level of performance [35]. Moreover, giving consumers ability to interact with the application using different types of multimedia increase their attention and engagement [36, 37]. Customer-focus is another indicator of a good application. It involves communicating with consumers to understand their needs and make changes accordingly to improve their navigation experience [28, 29, 37, 38-40]. The final indicator that distinguishes a good e-commerce application is the level of assurance. High level of assurance is determined by providing safe payment options, gaining security certificates and applications pass certification, presenting security, privacy, and copyright policies, providing easy to access contacts, and offering alternative supporting sources [37, 41-44]. Furthermore, the literature details the above discussed six indicators into measurable elements. In the guidelines of the literature, we discuss these elements and link them with the IS success model constructs to develop a conceptual framework.

*System Quality*- represents one of functional standards in e-commerce and consists of reliability, responsiveness and flexibility. It is important to deal with online purchasing through the Internet or mobile environments [45]. Due to the importance of system quality, which impacts significantly on using commercial websites, the system performance of the e-commerce applications should be designed and planned to be handled easily and conveniently for customers [16]. Additionally, system quality also measures the overall system performance according to participants, referred to here as consumers [18], so the e-commerce applications vary depending on consumer requirements of the electronic systems. Naturally, the e-commerce environment also varies based on actual consumers' use and environmental factors. The actual uses are voluntary, which means it requires the quality system to have high-performance standard to ensure an appropriate level of success [46]. Without performance quality of the system, it might be possible that customers' use of an M-Commerce system is affected negatively, and its perceived value will decrease below the expected level.

The requirements of system quality vary among previous studies. For instance, the Zhang and Dran [47] focused on eleven standard variables of system quality of e-commerce websites, while another study considers only five variables [48]. Thus, this study considers several variables used in previous studies related to e-commerce applications [27, 35, 45, 49]. Some previous studies interpreted system quality requirements using criteria such as usability, usefulness, reliability responsiveness, and flexibility [46]. Petter et al. [19] identified systems quality criteria as system being flexible, easy to use, easy to learning, satisfied of user requirements, accurate, sophisticated, integrated and customizable [19]. After considering examples of requirements used in previous studies, the requirements looked at in this study can be categorized into two main groups, appearance and organization. Both groups include number of indicators that determine actual and practical requirements of e-commerce smartphone applications. Thus, the System Quality was measured using four sub-constructs, namely: Attractive Appearance and Balancing (AP_AB), Color and Text usage (AP_CT), Application Planning and Consistency (OR_PC), and Navigation links (OR_NL). These four sub-constructs were measured using 20 items. Moreover, other three measurement items were linked directly to the system quality. Therefore, a total of 23 measurement items under system quality construct are identified based on the literature [27, 50-59].

Based on the previous discussion, the following system quality hypotheses are formulated.

*H1a*: The appearance balanced influences e-commerce smartphone application's system quality.
*H1b*: The colors and text influences e-commerce smartphone application's system quality.
*H2a*: The planning and consistency influences e-commerce smartphone application's system quality.
*H2b*: The Navigation links influences e-commerce smartphone application's system quality.

*Information Quality*- is frequently used as a fundamental standard in e-commerce contexts when applying the DeLone and McLean Model [60]. Information Quality can be defined as information used to assist consumers in deciding to participate in online purchasing, and it often has a positive link to success in e-commerce applications [45]. Thus, high quality information leads to successful implementation. The fundamental role of offering websites and e-commerce applications is to provide accurate and updated information about products and display it in an appropriate way to encourage online purchasing [61]. Moreover, DeLone and McLean indicated that information quality directly affects information uses and user satisfaction. The presentation and organization of information is also very important, especially as viewed through limited mobile screens [45, 62].

It appears that the variables used to measure information quality differ depending on the companies' conviction and level of maturity in presenting information to consumers. Additionally, different electronic sales channels (e.g. website – application – social media) help identify an appropriate amount of information presented through these channels, depending on the company's policies to increase the number of online sale transactions and customers. DeLone and McLean [18] mentioned that information quality can be limited to variables such as accuracy, relevance, currency, completeness and understandability. Furthermore, information quality can be measured in the following criteria: usability, understandability, availability, format, relevance and conciseness [19]. In relation to information quality and consumer requirements, these indicators may be placed in three main groups: content, assurance and interactions with the consumer. These three groups contain seven sub-constructs, namely: Updating Content and Relevant Information (CO_UI), containing four indicators; Accurate and Relevant Data (CO_AD), containing four indicators; Content Display (CO_CD), containing four indicators; Multimedia adoption (IN_MA), containing three indicators; Adaptability (IN_AD), containing four indicators; Customer advisor (IN_CD), containing five indicators; and Assurance (AS), containing thirteen indicators. Moreover, the indicators related directly to Information Quality are divided into five indicators, namely: preciseness, understandability, reliability, continuous update, and meeting needs. Therefore, a total of 42 indicators under Information Quality and related sub-constructs are identified based on the literature [36, 54-57, 63-67].

Based on the previous discussion, the following information quality hypotheses are formulated.

*H3a*: The Update Relevant Information influences e-commerce smartphone application's information quality.
*H3b*: The Accurate and Relevant Data influences e-commerce smartphone application's information quality.
*H3c*: The Content display influences e-commerce smartphone application's information quality.
*H4a*: The Multimedia adoption influences e-commerce smartphone application's information quality.
*H4b*: The Adaptability influences e-commerce smartphone application's information quality.
*H4c*: The Customer advisor influences e-commerce smartphone application's information quality.
*H5*: The Assurance influences e-commerce smartphone application's information quality.

*Service Quality*- represents a significant factor in e-commerce due to the limited connection between buyer and seller [68]. Service quality can be defined as a company's comprehensive customer support on e-commerce applications [16]. Consumer support by a company team member represents a key success factor in the DeLone and McLean model. Service quality represents many variables, such as reliability, responsiveness, tangibility and empathy [45]. Tools have been indicated to encourage customers finalizing their online shopping transactions, such as list of frequent asked questions, hotline help desk and online chat support to respond to customer inquiries in the shortest possible time [5, 69]. Furthermore, there was a focus on the importance of completing online purchasing using different promotions maintaining customer contact through electronic communication [45]. Rogers [70] emphasized on the importance of communication relationship tools. These tools encourage consumers to continue online purchasing through the same channel in the future [45]. Thus, indicators that are related to service quality can be organized in four sub-constructs of Customer Focus, namely: Mobile-loyalty Building (CF_MB), containing three indicators; Customer Chat and feedback (CF_CC), containing three indicators; Help and technical support (CF_HT), containing four indicators; and Credibility and Reliability Build (CF_CB), containing six indicators. Moreover, the indicators that directly linked to Service Quality are divided into Six indicators, namely: problem solving, help and support, security and privacy, answering questions, individual attention and understanding specific needs. Therefore, a total of a total of 22 indicators under service quality are identified [16, 25, 45, 51, 66, 71-73].

Based on the previous discussion, the following information quality hypotheses are formulated.

*H6a*: The M-loyalty Building influences e-commerce smartphone application's service quality.
*H6b*: The Customer Chat influences e-commerce smartphone application's service quality.
*H6c*: The Help and Technical Support influences e-commerce smartphone application's service quality.
*H6d*: The Credibility Build influences e-commerce smartphone application's service quality.

*User Satisfaction*- The appropriate use of electronic systems should have minimum requirements of quality that fulfil a required level of customer satisfaction. It is difficult to deny a system's appropriateness for consumers if their satisfaction is high [60]. Wu and Wang [60] purported that user satisfaction has a causal relationship to use of a system, though the opposite is not true. Thus, using the system refers to the temporal relationship with user satisfaction rather than a causal relationship [60]. The track used in this study is a one-directional relationship from user satisfaction to intention to use rather than the two-way relationship in the original DeLone and McLean model. The user satisfaction measurement represents the construct that depends on three quality constructs that focus on designing the e-commerce applications and assess their usability for consumers [74]. Furthermore, DeLone and McLean [18] indicated that consumer satisfaction is composed of three basic elements: the level of effective response, the time needed to reach the satisfaction level and the level of satisfaction after completing the purchase. Additionally, consumer satisfaction represents a main criterion used to distinguish the quality of goods and services, and high user satisfaction is reflected in future purchases rather than a switch to different online purchasing platforms [45]. A positive and direct relationship between the three quality dimensions and user satisfaction are highly correlated [45]. Thus, it can be considered that the user satisfaction intermediary relationship exists between quality requirements and consumer behavior requirements in e-commerce applications [75]. The User satisfaction construct contains three indicators, as follows. (1) You are satisfied with the e-commerce application system. (2) The e-commerce application system is of high quality. (3) The e-commerce application system has met users' expectations.

*Intention to use*- intention to use technical systems represents a positive reaction between the quality in an e-commerce application and user satisfaction. The intention to use also represents an inevitable result of user satisfaction when dealing with electronic systems. Therefore, user satisfaction plays a significant role in deciding a consumer's intention to use an e-commerce application periodically. Thus, the intention to use and user satisfaction determine the consumer behavior requirements with quality requirements in the online purchasing context [76]. Users who have previous experience dealing with merchants on the Internet exhibit less timidity to use e-purchasing than those without such experience [45]. Through the user satisfaction and intention to use indicators, it can evaluate the intention to use quality indicators and major functions that support e-commerce applications. Intention to use measures visits to mobile applications, navigation within the mobile applications, information retrieval and a transaction execution. Intention to Use can be measured by three indicators adopted from previous studies [77, 78]. The Intention to use indicators are as follows. (1) Intention to use any e-commerce application. (2) Reuse the e-commerce applications in the future. (3) Use the e-commerce applications frequently in the future.

*Net Benefit*- There are various definitions that determine the net benefits for the end user in the e-commerce context. The difference in net benefit definitions exist because the end user varies, and may be a designer, sponsor, or consumer; this fact shows that the net benefit seen from different views leads to a variation of responses when applying research tools [18]. The emphasis, in the current study, is on the consumers who are willing to purchase through e-commerce applications using their mobile devices. If consumers ultimately think that using e-commerce application will improve their experience with online purchasing, it would be helpful to raise the net benefit level of e-commerce applications [16, 60]. Through this research framework, the net benefits level focuses on three key aspects, namely, willingness to use, helpfulness and usefulness. The net benefit construct can contain four indicators, as follows. (1) The e-commerce product/service is good value for money. (2) The price of the e-commerce product/service is acceptable. (3) The time spent in the e-commerce application system is appropriate. (4) The e-commerce application system facilities will be extended to online shopping, thus increasing the consumer purchases.

*H7*: The e-commerce smartphone application's system quality significantly influences user satisfaction.
*H8*: The e-commerce smartphone application's information quality significantly influences user satisfaction.
*H9*: The e-commerce smartphone application's service quality significantly influences user satisfaction.
*H10*: The e-commerce smartphone application's system quality significantly influences intention to use.
*H11*: The e-commerce smartphone application's information quality significantly influences intention to use.
*H12*: The e-commerce smartphone application's service quality significantly influences intention to use.
*H13*: User satisfaction significantly influences the intention to use e-commerce smartphone application.
*H14*: User satisfaction significantly influences net benefit of e-commerce smartphone application.
*H15*: The intention to use e-commerce smartphone application significantly influences net benefit.

To sum up the review section, Table 1 presents the whole constructs summary, their relationships, number of measurement items and references. Figure 2 presents our modifications to the IS success model for the e-commerce smartphone applications with hypothesis.

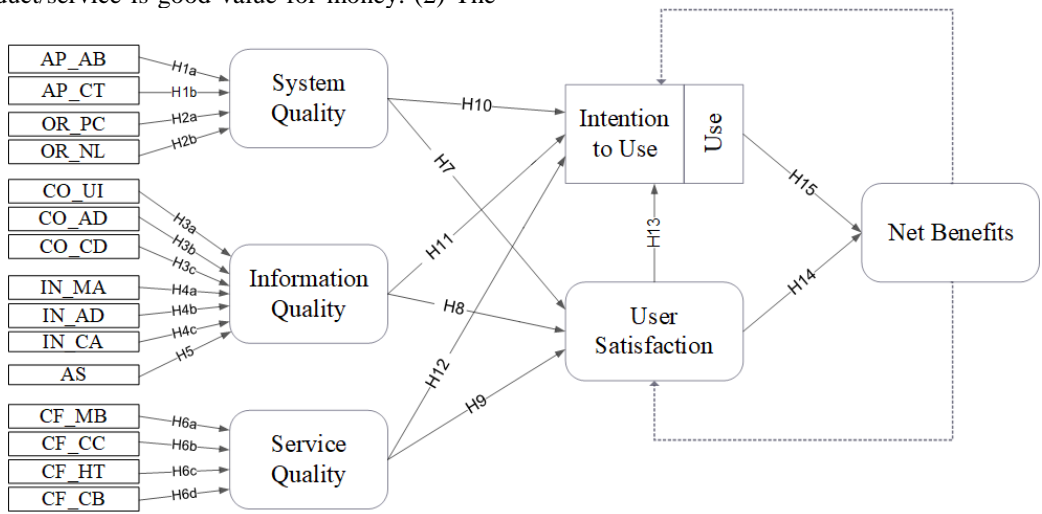

Figure 2: The modified IS success model for the e-commerce smartphone apps with hypothesis

Table 1: Summary of the new identified constructs with their measurement items and their relationships to the IS success model.

| Construct Code | Construct | Related to | Measurement items | References |
|---|---|---|---|---|
| AP_AB | Attractive Appearance and Balancing | System Quality | 5 | [27, 28, 29, 50, 51, 52, 58, 59] |
| AP_CT | Color and Text usage | System Quality | 6 | |
| OR_PC | Application Planning and Consistency | System Quality | 6 | [35, 50, 51, 54 66, 71] |
| OR_NL | Navigation links | System Quality | 3 | |
| SQ | System Quality | System Quality | 3 | [16, 45, 47, 60, 74, 75] |
| CO_UI | Updating Content and Relevant Information | Information Quality | 4 | [27, 30, 31, 32, 33, 34, 50, 51, 71, 79] |
| CO_AD | Accurate and Relevant Data | Information Quality | 4 | |
| CO_CD | Content Display | Information Quality | 4 | |
| IN_MA | Multimedia adoption | Information Quality | 3 | [36, 37, 50, 51, 54, 56, 57, 63, 64, 65, 67, 71, 80] |
| IN_AD | Adaptability | Information Quality | 4 | |
| IN_CD | Customer advisor | Information Quality | 5 | |
| AS | Assurance | Information Quality | 13 | [8, 37, 40, 41, 42, 43, 44, 50, 51, 66, 71, 81, 82, 83] |
| IQ | Information Quality | Information Quality | 6 | [16, 45, 60, 74, 75, 84, 85] |
| CF_MB | M-loyalty Building | Service Quality | 3 | [4, 9, 28, 29, 37, 38, 39, 40, 41, 51, 66, 71, 72, 73, 82, 83] |
| CF_CC | Customer Chat and feedback | Service Quality | 3 | |
| CF_HT | Help and technical support | Service Quality | 4 | |
| CF_CB | Credibility and Reliability Build | Service Quality | 6 | |
| SQU | Service Quality | Service Quality | 6 | [16, 45, 60, 63, 75, 84, 85] |
| US | User Satisfaction | Behavior Requirements | 3 | [10, 12, 16, 45, 58, 60, 63, 73, 75, 85] |
| IU | Intention to Use | Behavior Requirements | 3 | |
| NB | Net Benefits | Behavior Requirements | 4 | |

## 3. METHODOLOGY

The current study aims to define specifications of successful e-commerce smartphone applications for GCC consumers. To achieve this goal, based on the IS Success Model, a conceptual framework was developed. A survey instrument was designed grounded on the literature and the IS success model to examine the hypotheses of the developed model. The survey starts with two filtering questions to ensure that the participants are GCC citizens/residents and have had experience purchasing online using smartphones. This is followed by questions to collect participant demographic information and background. The main section of the survey contains 97 measurement items to be evaluated using a five-point Likert scale ranging from 1 (strongly agree) to 5 (strongly disagree). The survey, then, went through several steps to ensure survey validity. Arbitration process was performed. Arbitrating questionnaire plays a significant role in ensuring the credibility of data. The main goal of arbitration is focusing on the questions' concepts and measuring their relevance to the research objective. Therefore, the arbitrator answered on the questionnaire shall be as follows (Measures – does not measure – to a certain extent) during reading of the survey questions. The questionnaire should also include a separate sheet of paper describing the study and its objectives to help the arbitrator make valid decisions [86]. In the current study, the questionnaire survey was arbitrated by several specialists in the same field. They were from five academics working in different universities in the GCC region. This step determined the appropriateness of the questions and whether they were placed correctly. The purpose of this tool is to identify landmarks of consumer needs in Arabic societies. The arbitrators were asked to evaluate the whole 97 list of measurement items and their relevance. The major changes in the raised concerns were related to expressions of the statements to avoid ambiguity. Therefore, the revised version of the survey considered all the raised concerns. Since Arabic is the mother language in the GCC countries and English is ordinarily used in businesses, the questions of the survey were made available in both languages' Arabic and English, so that the participant could select the language they prefer.

The target research population is the smartphone users of GCC countries. The selection of the sample was made from the targeted research population. The sample size was determined based on the following formula [87].

$$X = Z \ (C/100)^2 * r \ (100 - r)$$
$$n = N * x / ((N-1) * E2 + x)$$
$$E = Sqrt \ [(N-n) \ x / n \ (N-1)]$$

(N) represents the number of GCC population who use smartphones.
(Z(C/100)) is the critical value for confidence level C.
(r) is the responses fraction that are of interest to the research team.
(C) is the confidence level.
(n) is the output of sample size.
(E) Signifies the error margin, calculated as 7.5 %.

Table 2 demonstrates the number of populations of the GCC countries, smartphone users (the research population), confidence level, merging of accepted error and the sample size base on the above-mentioned formula. Therefore, the minimum number of sample size should not be less than 644.

Table 2: Calculated Sample sizes of the GCC Population

| Item | No./percentage |
|---|---|
| Number of populations | 40533672 |
| Smartphone users' rate | 73.2% |
| Smartphone users' number | 29,670,650 |
| Confidence level | 99 |
| Margin of accepted error | 5 |
| Minimum sample size | 644 |

*Source:* [88-90]

Since the smartphone users were the main target for this study, the survey was mainly made available online using a mobile friendly version of open source LimeSurvey hosted by Griffith University. The online invitation was sent to the most prominent possible friends and interested people through social networking sites and mobile applications, forums and mailing groups. The total number of responses reached about one thousand. About 20% percent of the forms were excluded because they were incomplete. Therefore, the total of the completed forms that were used for the analysis reached 803.

The collected data was mainly analyzed employing exploratory and confirmatory factor analyses to test the developed model hypotheses. The analysis is detailed in the following section.

## 4. ANALYSIS AND RESULTS

At the beginning of analysis stage, the data screening was conducted to ensure the missing data, assess normality, screen outliers, and check the standards and standard errors. Once the data is ready, the second stage of analysis was performed. It measured the components' scaling, which included the following tests: internal consistency, item-total correlations, Exploratory Factor Analysis (EFA) and Confirmatory Factor Analysis (CFA). The third stage of analysis focused on studying the relationships among the model constructs particularly, and included the Composite Reliability (CR), Average Variance Extracted (AVE), Standardized Regression Coefficient (SRC), Critical Ratio and P-value tests. The following sections details the analysis stage.

### 4.1. Profile of Respondents

Table 3 demonstrates the descriptive statistics of the current study respondents. The total number respondents are 803. More than a half of the responses (57%) were received from male respondents while the rest (43%) were received from female respondents. They represent three Arabian Gulf countries: Saudi Arabia (KSA), United Arab Emirates (UAE) and Qatar; 48%, 31% and 21% respectively. Slightly more than two-thirds (68%) represent the age group 26-40, 13% is younger generation in the age group 18-25 and the rest 19% are in the age group of 41 and older. The vast majority of the respondents have either graduate or post-graduate qualifications: bachelor 42%, master 39%, and PhD 11% whereas the others have diploma 5% or high school 3%. When comes to the experience of purchasing online, almost all the respondents have had experience buying online. Respondents buy online every week 12%, more than one time a month 40%, once a month 24%, once every three months 13% and one time every three months 9%. Obviously, over two-third conduct online purchasing in a monthly basis.

Table 3: Respondents' Descriptive Statistics

| Info | Variable | Frequency (N) | Percentage (%) |
|---|---|---|---|
| All participants | | 803 | 100% |
| Gender | Male | 461 | 57% |
| | Female | 342 | 43% |
| Country | KSA | 386 | 48% |
| | UAE | 246 | 31% |
| | Qatar | 171 | 21% |
| Age group | 18-25 | 103 | 13% |
| | 26-30 | 203 | 25% |
| | 31-35 | 187 | 23% |
| | 36-40 | 159 | 20% |
| | 41-45 | 93 | 12% |
| | 45+ | 58 | 7% |
| Education Level | High School | 23 | 3% |
| | Diploma | 43 | 5% |
| | Bachelor | 334 | 42% |
| | Master | 317 | 39% |
| | PhD | 86 | 11% |
| Experience in purchasing online | never | 17 | 2% |
| | once every six months | 74 | 9% |
| | once every 3 months | 106 | 13% |
| | once a month | 189 | 24% |
| | 1+ every month | 319 | 40% |
| | every week | 98 | 12% |

### 4.2. Exploratory Factor Analysis (EFA)

EFA is a statistical procedure that is used to ensure and extract the main components and constructs with large number of variables. It is used to identify the relationships between items inside an individual construct and between the constructs in the study [91]. In this study, Principal Components Analysis (PCA) was used to extract the factors through several criteria such as Latent Root Criterion, Catell's Scree Test, the *a priori* criterion, and the Percentage of Variance Criterion [92].

Depending on the Eigen value, the Scree Test, the *a priori* criterion and the Percentage of Variance Criterion test; the structures that focus on identifying the basic indicators of the acceptance of e-commerce smartphone applications were determined. Table 4 summarizes the indicators proposed as a solution for implementing the IS success model. The percentage of cumulative variance extracted, about 68%, is satisfactory for solutions in the field of social sciences [92]. Seventeen indicators were eliminated from all indicators of this study because their values in EFA were below the acceptable loaded level. Finally, the Cronbach's alpha coefficients for all the indicators were high and they ranged between 0.617 and 0.911, which led to satisfactory internal consistency between the indicators no less than 0.60 [93].

Table 4: Summaries of EFA Results of all Samples together

| Construct | Indicators Removed | Factors Extracted | KMO | Cronbach's Alpha | Cumulative Variance | No. of Item | Description |
|---|---|---|---|---|---|---|---|
| System Quality constructs | 5 | 4 | 0.840 | 0.832 | 68.121 | 15 | It includes the constructs of Appearance (AP) and Organization (OR) indicators. |
| Information Quality constructs | 6 | 7 | 0.887 | 0.908 | 68.188 | 30 | It includes the constructs of Content (CO), Interaction (IN), and Assurance (AS) indicators. |
| Service Quality constructs | 4 | 4 | 0.888 | 0.889 | 68.141 | 12 | It includes the construct of Customer-focus (CF) indicators. |
| Other constructs | 7 | 6 | 0.729 | 0.701 | 68.142 | 17 | It includes the constructs of System Quality (SQ), Information Quality (IQ), Service Quality (SQ), Intention to use (IU), User Satisfaction (US), and Net Benefit (NB). |

At this stage, new subgroups emerged based on the correlations between the indicators. All the measurement items that were considered at this stage had a value of 0.5 or more for the factor loading. Some measurement items were found to have weak results and they did not belong to any new sub-constructs. Therefore, it was necessary to eliminate these items during the EFA process for future analysis. The Assurance construct showed cohesion among the indicators and it does not appear to have a division like the other constructs; therefore, this construct contains the largest number of items (12). New constructs were applied as AP (10 items, 2 constructs), OR (7 items, 2 constructs), CO (9 items, 3 constructs), IN (11 items, 3 constructs), AS (12 items, 1 construct), CF (16 items, 4 constructs), SQ (3 items, 1 construct), IQ (4 items, 1 construct), SQU (3 items, 1 construct), IU (3 items, 1 construct), US (3 items, 1 construct), NB (3 items, 1 construct).

### 4.3. Measurement Model Specification and Assessment Criteria

This approach determines the measurement model specifications and assesses their accuracy. The measurement model in the Confirmatory Factor Analysis (CFA) presents the relationships between the variables, which helps in measuring the constructs that cannot be measured directly [92]. The measurement model was developed by integrating the models of the CFA constructs. This model consists of three layers:
1. Constructs, which show the scaling factors.
2. The first level of the modified IS success model containing the three qualities constructs (system, information and service). The layer is named "system requirements".
3. The second level of the modified IS success model containing the User Satisfaction, Intention to Use, and Net Benefit constructs. The layer is called "consumer behavioural requirements".

CFA tests was utilized to assess the measurement model. The main test and recommended values of the CFA assessment of model fit standards are presented in Table 5 below.

Table 5: The recommended Model Fit Criteria

| No. | Estimate indices | References |
|---|---|---|
| 1. | $X^2/df < 3.0$ | [92, 94] |
| 2. | GFI, TLI, NFI, CFI and IFI $> 0.90$ | [92] |
| 3. | AGFI $> 0.80$ | [94-96] |
| 4. | RMR & RMSEA $< 0.08$ | [92] |
| 5. | Correlation Coefficients $< 0.850$ | [94, 97] |
| 6. | t-values Or $R^2$, Factor Loadings $> 0.5$ | [92] |
| 7. | The Critical Ratio $> 1.96$ | [95, 96] |

Adding to the previous tests, Composite Reliability (CR) and Average Variance Extracted (AVE) were performed, which are more accurate tests in the CFA stage. If the value of CR is high degree, the indicators will have high reliability in the constructs and theoretical framework [92, 98]. In this study, the acceptable value of CR is equal or greater than 0.6, and the AVE should be equal or more than 0.5 [99]. As well as the correlation between the constructs should be less than the value of square root of AVE in the same construct and this is evidence of the health of differentiation between the constructs in the model [100, 101].

The statistics in Table 5 indicates the relationship between the constructs in the research model. The presented results demonstrate that the correlation between the constructs in which the square root of the average values is greater than the other correlation constructs is in the same group. In addition, the AVE values were around 0.5. All the CR values ranged between 0.7324–0.9346. The largest value for the squared multiple correlations was 0.774, while the smallest value was 0.469. Furthermore, the AVE values ranged between 0.7915–0.4706. Thus, the discriminant validity of the employed measurement items is adequate for the current research model.

The developed hypotheses that connected the constructs with one another were tested. They were divided into four subgroups: the SQ hypotheses, the IQ hypotheses, the SQU hypotheses and the IS success hypotheses. To evaluate these hypotheses, the reliability of the measures, discriminant validity and convergent validity were tested. Fornell and Larcker [100] recommended three criteria to evaluate the Convergent validity:
1. The Factor Loading Value should be more than 0.50.
2. The Composite Reliability should be more than 0.6.
3. The Average Variance Extracted must be not less than 0.5.

The analysis result of the SQ constructs, which include the constructs in the AP and OR constructs, showed a strong link between their indicators and the SQ constructs. This shows the meaningful relationship between the constructs, and the correlation values between the constructs ranged from 0.168 to 0.283. The t-value, which is of critical value, should be what is used to demonstrate convergent validity results of the constructs. The t-value estimates the parameter from its notional value and its standard error. In this study, the t-value in the same constructs ranged between 5.507 and 3.491, which means it is in the measurement average [92]. Moreover, CO, IN, and AS, which are connected to IQ, have good correlation coefficient values. Those values ranged between 0.493 and 0.347, and they were all significant values, while the t-value results ranged between 8.285 and 6.170. In the CF constructs, which are associated with SQU, the correlation coefficient values ranged between 0.131 and 0.488 and all the values were significant. Their t-values ranged between 3.404 and 5.920. In the ISS constructs, the correlation values ranged between 0.213 and 0.472 and all the constructs were significant in these constructs. The t-values of the ISS constructs ranged between 4.255 and 8.273. This means all the constructs had a significant correlation and they met the minimum t-value, which is greater than 1.96. Additionally, the Cronbach's alpha values for all the constructs ranged between 0.617 and 0.893, and the composite reliability values ranged between 0.9346 and 0.6153. In addition, the squared multiple correlation values ranged between 0.774 and 0.469.

The AVE test values, see Table 6, ranged between 0.7915 and 0.5006, which indicates good internal consistency between the indicators in the model as recommended by Fornell and Larcker [100]. These results indicate that the measurement model of e-commerce smartphone applications has a large convergent validity. Furthermore, appropriate correlation coefficient result should be less than 0.850, which lead the scale validity of discriminant was sufficient [94]. The AVE is used to check the discriminant validity through comparing the square root of the AVE with each construct results in the same groups in the model. The result shows all constructs were closely related to same groups instead of other constructs [100].

Table 6 Correlation Matrix and Discriminant Validity of the Measurement Model

| The AVE of System Quality Constructs | | | | | | |
|---|---|---|---|---|---|---|
| Construct Code | AP_AB | AP_CR | OR_PC | OR_NL | Mean | SD |
| AP_AB | **0.710** | | | | 4.2994 | 0.6922 |
| AP_CR | 0.417 | **0.708** | | | 4.0881 | 0.8210 |
| OR_PC | 0.461 | 0.281 | **0.783** | | 4.4209 | 0.6085 |
| OR_NL | 0.397 | 0.259 | 0.676 | **0.826** | 4.6451 | 0.5415 |

| The AVE of Information Quality Constructs | | | | | | | | | |
|---|---|---|---|---|---|---|---|---|---|
| Construct Code | CO_UI | CO_AC | CO_CD | IN_MA | IN_AD | IN_CA | AS_ALL | Mean | SD |
| CO_UI | **0.728** | | | | | | | 4.4251 | 0.6551 |
| CO_AC | 0.573 | **0.772** | | | | | | 4.1606 | 0.7703 |
| CO_CD | 0.434 | 0.427 | **0.778** | | | | | 4.4938 | 0.7090 |
| IN_MA | 0.364 | 0.414 | 0.352 | **0.760** | | | | 4.2989 | 0.6946 |
| IN_AD | 0.588 | 0.407 | 0.393 | 0.633 | **0.728** | | | 4.4645 | 0.6448 |
| IN_CA | 0.333 | 0.466 | 0.281 | 0.628 | 0.578 | **0.760** | | 3.9288 | 0.7956 |
| AS_ALL | 0.373 | 0.326 | 0.302 | 0.364 | 0.490 | 0.379 | **0.768** | 4.6599 | 0.5948 |

| The AVE of Service Quality Constructs | | | | | | |
|---|---|---|---|---|---|---|
| Construct Code | CF_HT | CF_MB | CF_CC | CF_CB | Mean | SD |
| CF_HT | **0.890** | | | | 4.6613 | 0.5745 |
| CF_MB | 0.451 | **0.883** | | | 4.5160 | 0.7240 |
| CF_CC | 0.438 | 0.517 | **0.797** | | 4.3454 | 0.6693 |
| CF_CB | 0.581 | 0.584 | 0.577 | **0.781** | 4.6989 | 0.5439 |

| The AVE of IS success Constructs | | | | | | | | |
|---|---|---|---|---|---|---|---|---|
| Construct Code | US | SQ | IQ | SQU | IU | NB | Mean | SD |
| US | **0.832** | | | | | | 4.636 | 0.577 |
| SQ | 0.228 | **0.815** | | | | | 4.457 | 0.719 |
| IQ | 0.472 | 0.542 | **0.775** | | | | 4.554 | 0.631 |
| SQU | 0.428 | 0.092 | 0.179 | **0.773** | | | 4.447 | 0.683 |
| IU | 0.213 | 0.363 | 0.283 | 0.239 | **0.849** | | 4.102 | 0.744 |
| NB | 0.215 | 0.048 | 0.315 | 0.132 | 0.324 | **0.795** | 3.996 | 1.016 |

Furthermore, the goodness-of-fit tests were performed for each construct, see Table 7. The results show proportional model at recommended values, the relationship between model's constructs in all study samples, and the indicators have acceptable value in entire the study model.

Table 7  Model-Fit Indices of the Research Model Constructs

| The goodness-of-fit & Recommended Value | | System Quality | Information Quality | Service Quality | Other ISS constructs | The Result of the sample |
|---|---|---|---|---|---|---|
| X²/df | ≤ 3.00 | 3.58 | 4.112 | 3.402 | 2.904 | 3.100 |
| GFI | ≥ 0.9 | 0.957 | 0.873 | 0.958 | 0.990 | 0.945 |
| TLI | ≥ 0.9 | 0.936 | 0.833 | 0.94 | 0.985 | 0.924 |
| NFI | ≥ 0.9 | 0.918 | 0.816 | 0.936 | 0.970 | 0.910 |
| CFI | ≥ 0.9 | 0.948 | 0.853 | 0.954 | 0.990 | 0.936 |
| IFI | ≥ 0.9 | 0.948 | 0.854 | 0.954 | 0.990 | 0.937 |
| AGFI | ≥ 0.8 | 0.942 | 0.847 | 0.936 | 0.981 | 0.927 |
| RMR | ≤ 0.8 | 0.020 | 0.022 | 0.015 | 0.019 | 0.019 |
| RMSEA | ≤ 0.8 | 0.044 | 0.062 | 0.055 | 0.025 | 0.047 |

### 4.4. The hypotheses test

Based on the previous results, the basic assumptions relating to this model consisted of 24 hypotheses covering the entire proposed model. All the hypotheses and their relationships are presented in Table 8 including the standardized path coefficients, the t-values, and the $p$-values. Obviously, the statistical evidence supports all the proposed hypotheses of the modified IS success model for e-commerce smartphone apps.

Table 8: Path Coefficients, t-values, and $p$-values of the e-commerce smartphone application hypothesis

| The relationship for path | Standardized path coefficient | Critical ratio or (t-value) | $p$-value | Composite Reliability values (CR) (>0.6) | AVE (> 0.5) | Remarks |
|---|---|---|---|---|---|---|
| AP_ AB → SQ | 0.383 | 5.507 | 0.001** | 0.8302 | 0.5036 | Accepted |
| AP_ CR → SQ | 0.374 | 3.491 | 0.002** | 0.7779 | 0.5006 | Accepted |
| OR_ PC → SQ | 0.368 | 3.621 | 0.001** | 0.8629 | 0.6137 | Accepted |
| OR_ NL → SQ | 0.310 | 4.315 | 0.003** | 0.8097 | 0.6815 | Accepted |
| CO_ UI → IQ | 0.411 | 7.052 | 0.001** | 0.7699 | 0.5298 | Accepted |
| CO_ AC → IQ | 0.347 | 6.170 | 0.002** | 0.8101 | 0.5959 | Accepted |
| CO_CD → IQ | 0.353 | 6.217 | 0.001** | 0.7538 | 0.6049 | Accepted |
| IN_MA → IQ | 0.412 | 6.575 | 0.003** | 0.7324 | 0.5779 | Accepted |
| IN_AD → IQ | 0.493 | 8.285 | 0.004** | 0.8179 | 0.5295 | Accepted |
| IN_CA → IQ | 0.434 | 7.822 | 0.002** | 0.8718 | 0.5773 | Accepted |
| AS_ALL → IQ | 0.381 | 7.497 | 0.001** | 0.9346 | 0.5895 | Accepted |
| CF_HT → SQU | 0.581 | 5.920 | 0.04* | 0.8836 | 0.7915 | Accepted |
| CF_MB → SQU | 0.451 | 3.404 | 0.03* | 0.9136 | 0.7802 | Accepted |
| CF_CC → SQU | 0.438 | 3.545 | 0.009** | 0.8391 | 0.6351 | Accepted |
| CF_CB → SQU | 0.488 | 4.764 | 0.02* | 0.8620 | 0.6107 | Accepted |
| SQ → IU | 0.363 | 5.254 | 0.005** | 0.6153 | 0.6923 | Accepted |
| SQ → US | 0.328 | 4.743 | 0.001** | 0.7251 | 0.6218 | Accepted |
| IQ → IU | 0.383 | 6.581 | 0.003** | 0.6702 | 0.665 | Accepted |
| IQ → US | 0.472 | 8.273 | 0.001** | 0.7123 | 0.6827 | Accepted |
| SQU → IU | 0.339 | 7.481 | 0.002** | 0.6775 | 0.6003 | Accepted |
| SQU → US | 0.428 | 5.508 | 0.003** | 0.6175 | 0.6823 | Accepted |
| US → IU | 0.313 | 4.255 | 0.009** | 0.6741 | 0.597 | Accepted |
| IU → NB | 0.324 | 6.472 | 0.005** | 0.6500 | 0.7203 | Accepted |
| US → NB | 0.315 | 5.307 | 0.006** | 0.6312 | 0.6327 | Accepted |

** Significant at $p < 0.01$, * Significant at $p < 0.05$

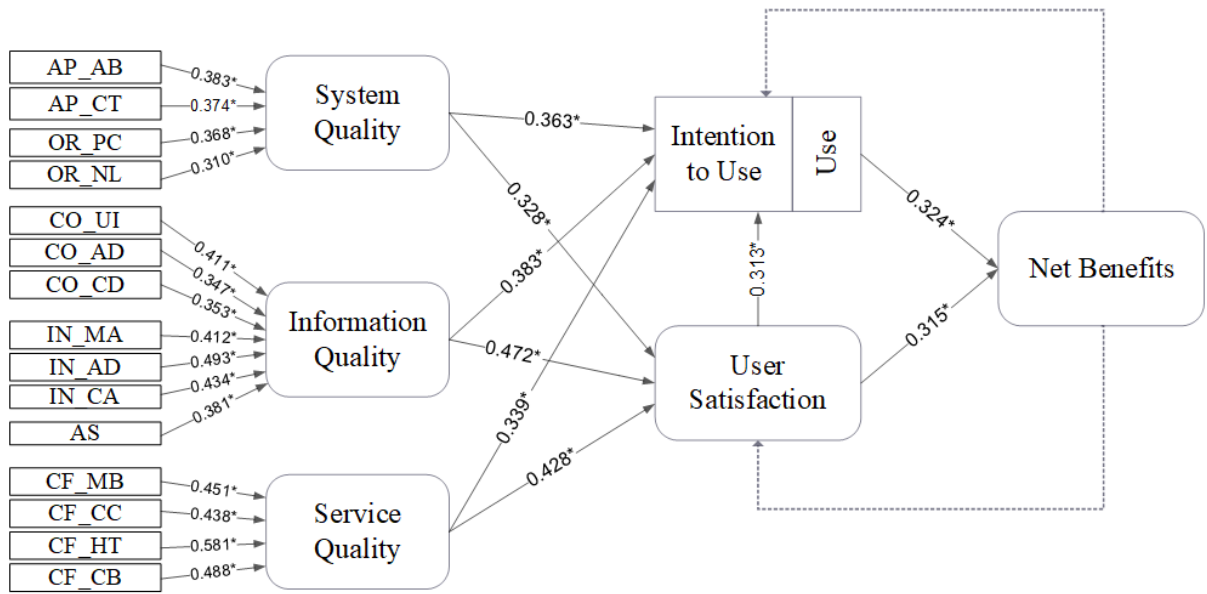

Figure 3: The modified IS success model for e-commerce smartphone apps with the tested hypotheses

## 5. DISCUSSION

The Main question of the current study focuses on the functions that enhance the use of e-commerce smartphone applications in the GCC society. The indicators of the fifteen sub-constructs of our conceptual framework were identified as functions for e-commerce smartphone applications. These functions were organized around three categories: technological (with eleven functions), organizational (with four functions) and social (with five functions).

The requirements are divided among the functions that have similar tasks and goals in order to reduce the number of functions that would need to be designed in the following stages. As these requirements are basically associated with SQ, IQ, SQU, they are divided up according to the similarity criteria of input, processing and output. The 20 functions belonging to the six main groups of sub-constructs which are AP, CO, OR, IN, CF and AS. These functions are transferred from the indicators to functions after identifying the main purpose of each indicator is defined; thus, they are converted to actions. These functions are divided into public functions and special functions. Public functions affect all the application screens, while special functions affect only specific screens.

Assessment of consumer behaviour characteristics in the study sample represents an important element in the successful use of e-commerce smartphone applications; this concentrates on determining the indicators that affect a consumer's decision to accept an online purchase or not. The technical application requirements focus on increasing the level of acceptance among consumers, but consumer behaviour indicators are concerned with measuring the consumer level of acceptance for User Satisfaction, the Intention to Use in the future and the Net Benefits in dealing with e-commerce smartphone applications. The causal relationships between these indicators help to determine the requirements of each construct of the consumer behaviour perspective. The IU indicators focus on measuring the causal relationship in the user's Intention to Use, which US indicators focus on the causal relationship between the US and IU which is the US cause IU. The US indicators determine the received information and methods that are used in smartphones' commercial application systems. Net Benefit reflects the positive advantages for the individual, society and operators or owners of technical application companies. The level of trust among consumers can be measured through the commercial applications by increasing the level of US and IU Use regularly in the future.

## 6. CONCLUSION

Online business has grown exponentially throughout the world. This expansion of online business in the GCC region was further progressed through smartphone applications. This research studied the impact of various factors that influence the success of the smartphone applications and in turn causing the expansion of online business. The research studied these factors using IS Success model and developed a theoretical model, which was further tested using a survey responded by 803 participants. Exploratory and confirmatory analysis shows that various factors pertaining to System Quality, Information Quality, Service Quality, Intention to Use and User Satisfaction influence the effectiveness of smartphone application. All the factors considered in the theoretical model proved to have sufficient impact on online business expansion. In specific, factors like Technical Support, Credibility Build (from Service Quality construct) and Adaptability (from Information Quality) have higher standardized path coefficients among all the factors studied. The proposed conceptual model, therefore, should serve as a guide for commercial organisations in the GCC region when developing their smartphone applications.

# REFERENCES

[1] Laudon, K. & Traver, C. (2015). *E-commerce: business, technology, society.* Pearson Prentice Hall: New Jersey

[2] Statista (2018). B2C e-commerce sales worldwide from 2012 to 2018 (in billion U.S. dollars) Retrieved February 22, 2019, from http://www.statista.com/statistics/261245/b2c-e-commerce-sales-worldwide/

[3] e-Marketer. (2018). *Smartphones, Mobile Commerce Roundup,* Retrieved February 21, 2019, from https://www.emarketer.com/public_media/docs/eMarketer_Mobile_Commerce_Roundup.pdf

[4] AlGhamdi, R. (2014). Diffusion of the Adoption of Online Retailing in Saudi Arabia. (PhD), Griffith University, Brisbane, Australia.

[5] Aladwani, A. M. (2003). Key Internet characteristics and e-commerce issues in Arab countries. *Information Technology & People*, *16*(1), 9-20.

[6] Al-Solbi, A., & Mayhew, P. J. (2005). Measuring e-readiness assessment in Saudi organisations preliminary results from a survey study. *From e-government to m-government*, 467-475.

[7] Alrawi, K. W., & Sabry, K. A. (2009). E-commerce evolution: A Gulf region review. *International Journal of Business Information Systems*, *4*(5), 509-526.

[8] Alshehri, M., Drew, S., Alhussain, T., & Alghamdi, R. (2012). The Effects of Website Quality on Adoption of E-Government Service: An Empirical Study Applying UTAUT Model Using SEM. in J Lamp (ed.), *23rd Australasian Conference On Information Systems (ACIS 2012)*, Melbourne, Australia.

[9] AlGhamdi, R., Nguyen, J., Nguyen, A., & Drew, S. (2012). Factors influencing e-commerce adoption by retailers in Saudi Arabia: A quantitative analysis. *International Journal of Electronic Commerce Studies*, *3*(1), 83-100.

[10] Bahaddad, A. A., AlGhamdi, R., & Houghton, L. (2012). To What Extent Would E-mall Enable SMEs to Adopt E-Commerce? *International Journal of Business and Management, 7*(22), 123-132.

[11] AlGhamdi, R., Nguyen, A. & Jones, V. (2013), A Study of Influential Factors in the Adoption and Diffusion of B2C E-Commerce, In: *International Journal of Advanced Computer Science and Applications (IJACSA) 4*(1). 89-94.

[12] Bahaddad, A. A., Drew, S., Houghtoni, L., & Alfarraj, O. A. (2015). Factors attracting online consumers to choose e-Malls for e-procurement in Saudi Arabia. *Enterprise Information Systems*, *12*(7), 856-887.

[13] Le-Nguyen, K., & Guo, Y. (2016). Choosing e-commerce strategies: a case study of eBay. vn partnership. *Journal of Information Technology Teaching Cases*, *6*(1), 1-14.

[14] AlGhamdi, R., Alfarraj, O. A., & Bahaddad, A. A. (2015). How Retailers at different Stages of E-Commerce Maturity Evaluate Their Entry to E-Commerce Activities? *arXiv preprint arXiv:1503.05172*.

[15] Chen, C. W. D., & Cheng, C. Y. J. (2009). Understanding consumer intention in online shopping: a respecification and validation of the DeLone and McLean model. *Behaviour & Information Technology*, *28*(4), 335-345.

[16] Niranjanamurthy, M., & Kavyashree, N. (2013). Analysis of E-Commerce and M-Commerce: Advantages, Limitations and Security issues. *International Journal of Advanced Research in Computer and Communication Engineering*, *2*(6). 2360-2370

[17] Delone, W. H., & McLean, E. R. (2003). The DeLone and McLean model of information systems success: a ten-year update. *Journal of management information systems*, *19*(4), 9-30.

[18] Petter, S., DeLone, W., & McLean, E. (2008). Measuring information systems success: models, dimensions, measures, and interrelationships. *European Journal of Information Systems, 17*(3), 236-263.

[19] Gable, G. G., Sedera, D., & Chan, T. (2008). Re-conceptualizing information system success: The IS-impact measurement model. *Journal of the association for information systems*, *9*(7), 377-408.

[20] McGill, T., Hobbs, V., & Klobas, J. (2003). User developed applications and information systems success: A test of DeLone and McLean's model. *Information Resources Management Journal (IRMJ), 16*(1), 24-45.

[21] Mahatanankoon, P., Wen, H. J., & Lim, B. (2005). Consumer-based M-Commerce: exploring consumer perception of mobile applications. *Computer standards & interfaces*, *27*(4), 347-357.

[22] Petter, S., DeLone, W., & McLean, E. R. (2013). Information Systems Success: the quest for the independent variables. *Journal of Management Information Systems*, *29*(4), 7-62.

[23] Ho, R., Huang, L., Huang, S., Lee, T., Rosten, A., & Tang, C. S. (2009). An approach to develop effective customer loyalty programs: The VIP program at T&T Supermarkets Inc. *Managing Service Quality: An International Journal*, *19*(6), 702-720.

[24] Varnali, K., & Toker, A. (2010). Mobile marketing research: The-state-of-the-art. *International Journal of Information Management*, *30*(2), 144-151.

[25] Hossain, M. Y., & Hossain, M. (2011). E-Service Quality and Consumer Loyalty: A study on Consumer Electronic Retail Industry, *Master thesis*, Umeå University. Avialable from http://www.diva-portal.org/smash/get/diva2:524922/FULLTEXT02.pdf

[26] Hasan, L. & Abuelrub, E. (2011). Assessing the quality of web sites. *Applied Computing and Informatics*. 9(1), 11-29.

[27] Zahra, S., Khalid, A., & Javed, A. (2013). An Efficient and Effective New Generation Objective Quality Model for Mobile Applications. *International Journal of Modern Education and Computer Science*, *5*(4), 36-44.

[28] Salameh, A., Hassan, S., Alekam, J., & Alkafagi, A., (2015). Assessing the Effect of Service Quality and Information Quality on Customers' Overall Perceived Service Quality in M-Commerce. *Australian Journal of Basic and Applied Sciences*, *9*(13), 146-153.

[29] Sohaib, O., & Kang, K. (2013). The Importance of Web Accessibility in Business to-Consumer (B2C) Websites. In: *22nd Australasian Software Engineering Conference (ASWEC 2013)*.

[30] AlGhamdi, R., Nguyen, A., Nguyen, J., & Drew, S. (2011). Factors Influencing Saudi Customers' Decisions to Purchase from Online Retailers in Saudi Arabia: A Quantitative Analysis. In: *IADIS International Conference e-Commerce*, Rome, Italy.

[31] Mansour, Y. (2014). Regulating payments for M-Content: The positive impact of the deregulation. *International Review of Law*, 9-20.
[32] AlGhamdi, R., Drew, S. & Al-Ghaith, W. (2011). Factors Influencing E-Commerce Adoption by Retailers in Saudi Arabia: A Qualitative Analysis, *The Electronic Journal of Information Systems in Developing Countries*, 47(7), 1-23.
[33] Noh, M. J., & Lee, K. T. (2016). An analysis of the relationship between quality and user acceptance in smartphone apps. *Information Systems and E-Business Management*, 14(2), 273-291.
[34] Page, T. (2013). Usability of text input interfaces in smartphones. *Journal of Design Research*, 11(1), 39-56.
[35] Li, H. & Suomi, R. (2009). A proposed scale for measuring e-service quality. *International Journal of u-and e-Service, Science and Technology,2*(1), 1-10
[36] Pousttchi, K., Tilson, D., Lyytinen, K., & Hufenbach, Y. (2015). Introduction to the Special Issue on Mobile Commerce: Mobile Commerce Research Yesterday, Today, Tomorrow—What Remains to Be Done? *International Journal of Electronic Commerce*, 19(4), 1-20.
[37] Bahadad, A., AlGhamdi, R. & Alkhalaf, S. (2014), Adoption Factors for e-Malls in the SME Sector in Saudi Arabia, In: *International Journal of Computer Science and Information Technologies (IJCSIT)*, 5(4), 5835-5856.
[38] Suki, N. M., & Suki, N. M. (2014). Mobile social networking service users' trust and loyalty. *Mobile Electronic Commerce: Foundations, Development, and Applications*, 89-106.
[39] Lu, M. T., Hu, S. K., Huang, L. H., & Tzeng, G. H. (2015). Evaluating the implementation of business-to-business m-commerce by SMEs based on a new hybrid MADM model. *Management Decision*, 53(2), 290-317.
[40] Wahab, S., Zahari, A. S. M., Al Momani, K., & Nor, N. A. M. (2011). The influence of perceived privacy on customer loyalty in mobile phone services: An Empirical Research in Jordan. *International Journal of Computer Science Issues (IJCSI)*, 8(2), 45-52.
[41] Nelson, P. (2014). How to turn on Android encryption today (no waiting necessary Retrieved December 17, 2015 from http://www.greenbot.com/article/2145380/why-and-how-to-encrypt-your-android-device. html
[42] Mohamed, I., & Patel, D. (2015, April). Android vs iOS Security: A Comparative Study. In: *Information Technology-New Generations (ITNG), 2015 12th International Conference on,* 725-730. IEEE.
[43] Zheng, M., Xue, H., Zhang, Y., Wei, T., & Lui, J. C. (2015). Enpublic Apps: Security Threats Using iOS Enterprise and Developer Certificates. In: *Proceedings of the 10th ACM Symposium on Information, Computer and Communications Security,* 463-474. ACM.
[44] Kuan, H. H., Bock, G. W., & Vathanophas, V. (2008). Comparing the effects of website quality on customer initial purchase and continued purchase at e-commerce websites. *Behaviour & Information Technology*, 27(1), 3-16.
[45] Delone, W. H., & Mclean, E. R. (2004). Measuring e-commerce success: applying the DeLone & McLean information systems success model. *International Journal of Electronic Commerce,* 9(1), 31-47.
[46] Zhang, P., & Von Dran, G. M. (2000). Satisfiers and dissatisfiers: A two-factor model for website design and evaluation. *Journal of the American society for information science*, 51(14), 1253-1268.
[47] Kim, S., & Stoel, L. (2004). Apparel retailers: website quality dimensions and satisfaction. *Journal of Retailing and Consumer Services*, 11(2), 109-117.
[48] Alotaibi, M. (2013). E-Commerce Adoption in Saudi Arabia: An Assessment of International, Regional and Domestic Web Presence. *I.J. Information Technology and Computer Science,* 5(2), 42-56.
[49] Barnes, S. J. & Vidgen, R. T., (2002). An integrative approach to the assessment of e-commerce quality. *Journal of Electronic Commerce Research, 3*(3), 114-127.
[50] Basu, A. (2003). Context-driven assessment of commercial web sites. In: *the 36th Annual Hawaii International Conference on System Sciences* Hawaii, USA: IEEE.
[51] Tan, F. B. & Tung, L. L. (2003). Exploring website evaluation criteria using the repertory grid technique: A web designers' perspective. In: *the second annual workshop on HCI research in MIS.* Seattle, WA, Australia: AIS; AUT University
[52] Wenham, D. & Zaphiris, P. (2003). User interface evaluation methods for internet banking web sites: a review, evaluation and case study. *Human-Computer Interaction, Theory and Practice,* 721-725
[53] Mich, L., Franch M., & Gaio, L. (2003). Evaluating and designing Web site quality. *Multimedia, IEEE, 10*(1), 34-43.
[54] Achour, H. & Bensedrine, N. (2005). An evaluation of internet banking and online brokerage in Tunisia. In: *1st International Conference on E-Business and E-learning.* Amman, Jordan
[55] Signore O., A. (2005). Comprehensive model for web sites quality. *Paper presented at the 7th IEEE International Symposium on Web Site Evolution*, Pisa, Italy: IEEE.
*[56]* Awamleh, R. & Fernandes, C., (2005). Internet Banking: An empirical investigation into the extent of adoption by banks and the determinants of customer satisfaction in the United Arab Emirates. *Journal of Internet Banking and Commerce, 10*(1), 1-12.
[57] Chen, S. Y. & Macredie, R. D. (2005). The assessment of usability of electronic shopping: a heuristic evaluation. *International Journal of Information Management, 25*(6), 516-532
[58] ISO, (2008). *Ergonomics of human-system interaction -- Part 151: Guidance on World Wide Web user interfaces*, ISO 9241-151:2008. Retrieved April 30, 2015, from http://www.iso.org/iso/iso_catalogue/ catalogue_tc/catalogue_detail. htm?csnumber=37031
[59] Wu, J. H., & Wang, Y. M. (2006). Measuring KMS success: A respecification of the DeLone and McLean's model. *Information & Management*, 43(6), 728-739.
[60] Cao, M., Zhang, Q., & Seydel, J. (2005). B2C e-commerce web site quality: an empirical examination. *Industrial Management & Data Systems*, 105(5), 645-661.
[61] Kim, S. C., Yoon, D., & Han, E. K. (2016). Antecedents of mobile app usage among smartphone users. *Journal of marketing communications*, 22(6), 653-670.
[62] Rigas, D. I. & Memery, D. (2002). Utilising audio-visual stimuli in interactive information systems: a two-domain investigation on auditory metaphors. In: *International Conference on Information Technology: Coding and Computing,* Bradford, UK.
[63] Rigas, D. & Alty, J. (2005). The rising pitch metaphor: an empirical study. *International Journal of Human-Computer Studies, 62*(1), 1-20.

[64] Rigas D. & Stergiou, A. (2007). An Empirical Approach to Audio-Visual Guided Electronic Commerce. *WSEAS Transactions on Computer Research, 2*(2), 177-182.
[65] Lin, F., Huarng, K., Chen, Y., & Lin, S. (2004). Quality evaluation of web services. In: *IEEE International Conference on E-Commerce Technology for Dynamic E-Business.* Beijing, China
[66] Martins, J., Costa, C., Oliveira, T., Gonçalves, R., & Branco, F. (2019). How smartphone advertising influences consumers' purchase intention. *Journal of Business Research*, *94*, 378-387.
[67] Ahn, Y. S., Park, S. Y., Yoo, S. B., & Bae, H. Y. (2004). Extension of Geography Markup Language (GML) for mobile and location-based applications. In *International Conference on Computational Science and Its Applications,* 1079-1088. Springer, Berlin, Heidelberg.
[68] Palmer, J. W. (2002). Web site usability, design, and performance metrics. *Information systems research*, *13*(2), 151-167.
[69] Rogers, E. M. (2003). *Diffusion of innovations* (5th ed.). New york: Simon and Schuster.
[70] Fitzpatrick, S. (2000). *Everyday Stalinism: ordinary life in extraordinary times: Soviet Russia in the 1930s*. Oxford University Press, USA.
[71] Chung, W. & Paynter, J. (2002). An evaluation of Internet banking in New Zealand. In: *the 35th Annual Hawaii International Conference on System Sciences (HICSS)* Hawaii, USA: IEEE.
[72] Hajli, N., Sims, J., Zadeh, A. H., & Richard, M. O. (2017). A social commerce investigation of the role of trust in a social networking site on purchase intentions. *Journal of Business Research*, *71(1)*, 133-141.
[73] Rai, A., Lang, S. S., & Welker, R. B. (2002). Assessing the validity of IS success models: An empirical test and theoretical analysis. *Information systems research*, *13*(1), 50-69.
[74] Zviran, M., Glezer, C., & Avni, I. (2006). User satisfaction from commercial web sites: The effect of design and use. *Information & management*, *43*(2), 157-178.
[75] Anderson, R. E., & Srinivasan, S. S. (2003). E-satisfaction and e-loyalty: a contingency framework. *Psychology and Marketing*, *20*(2), 123-138.
[76] Lee, I., Choi, B., Kim, J., & Hong, S. J. (2007). Culture-technology fit: Effects of cultural characteristics on the post-adoption beliefs of mobile Internet users. *International Journal of Electronic Commerce*, *11*(4), 11-51.
[77] Wang, Y. S. (2008). Assessing e-commerce systems success: a re-specification and validation of the DeLone and McLean model of IS success. *Information Systems Journal*, *18*(5), 529-557.
[78] Talib, F., & Faisal, M. (2015). An analysis of the barriers to the proliferation of M-commerce in Qatar. *Journal of Systems and Information Technology*, *17*(1), 54-81.
[79] Yoo, B. & Donthu, N. (2001). Developing a scale to measure the perceived quality of an Internet shopping site (SITEQUAL). *Quarterly Journal of Electronic Commerce, 2*(1), 31-46
[80] AlMamari, M. O. H. A. M. M. E. D. (2007). Mobile commerce development in Oman. *Unpublished master's thesis,* The University of Sheffield. Available from http://dagda. shef. ac. uk/dissertations/2006-07/External/Al-MamariMohammed MScIS. pdf.
[81] Joubert, J., & Belle, J. P. V. (2009). The importance of trust and risk in M-Commerce: A South African perspective. *PACIS 2009 Proceedings*, 96-110.
[82] Khan, H., Talib, F., & Faisal, M. N. (2015). An analysis of the barriers to the proliferation of M-commerce in Qatar: A relationship modeling approach. *Journal of Systems and Information Technology*, *17*(1), 54-81.
[83] Seddon, P. B. (1997). A respecification and extension of the DeLone and McLean model of IS success. *Information systems research*, *8*(3), 240-253.
[84] Brynjolfsson, E., Hitt, L. M., & Yang, S. (2002). Intangible assets: Computers and organizational capital. *Brookings papers on economic activity, 2002*(1), 137-198.
[85] AlGhamdi, R., Nguyen, A. & Jones, V. (2013), Wheel of B2C E-commerce Development in Saudi Arabia, in: *Advances in Intelligent Systems and Computing,* Edited by: J.-H. Kim, E. T. Matson, H. Myung and P. Xu, pp. 1047-1055, Springer Berlin Heidelberg.
[86] Walliman, N. (2017). *Research methods: The basics*. 2nd Ed, Routledge, London.
[87] Raosoft. (2004). *Sample size calculator.* Retrieved from http://www.raosoft.com/samplesize.html
[88] QSAa (2015), Ministry of Development Planning and statistics – Statistics Sector, Retrieved September 6, 2015 from http://www.qsa.gov.qa/Eng/index.htm
[89] CDSI (2017), Statistical Summary of Teaching Adm. & Tec. Staff By Agency-2010, Retrieved March 11, 2017 from ttps://www.stats.gov.sa/sites/all/modules/pubdlcnt/pubdlcnt.php?file=https://www.stats.gov.sa/ sites/default/files/estm_pop_ar2017.xlsx&nid=12211
[90] UAE STATISTICS (2016). Population Estimates 2006 – 2010, Retrieved November 6, 2015 from http://www.uaestatistics.gov.ae/EnglishHome/ReportDetailsEnglish/tabid/121/Default.aspx?ItemId=1914&PTID=104&MenuId=1
[91] Norris, M., & Lecavalier, L. (2010). Evaluating the use of exploratory factor analysis in developmental disability psychological research. *Journal of autism and developmental disorders*, *40*(1), 8-20.
[92] Hair, J. F., Black, W., Babin, B., & Anderson, R. (2010). *Multivariate data analysis: a global perspective* (7th ed.). New Jersey: Pearson.
[93] Pallant, J. (2005). *SPSS Survival Manual: A Step by Step Guide to Data Analysis using SPSS for Windows (Version 12)*. Berkshire: Open University Press.
[94] Kline, R. B. (2005). *Principles and Practice of Structural Equation Modeling* (2nd ed.). New York: Guilford Press.
[95] Bhattacherjee, A. (2001). Understanding information systems continuance: an expectation-confirmation model. *MIS quarterly*, 351-370.
[96] Chin, W. W., & Todd, P. A. (1995). On the use, usefulness, and ease of use of structural equation modeling in MIS research: a note of caution. *MIS quarterly*, 237-246.
[97] Tabachnick, B. G., & Fidell, L. S. (2007). *Using Multivariate Statistics*. Boston: Pearson Education, Inc.
[98] Kouftheros, X. A. (1999). Testing a model of pull production: a paradigm for manufacturing research using structural equation modeling. *Journal of Operations Management*, *17*(4), 467-488.
[99] Bagozzi, R. P. (1992). The self-regulation of attitudes, intentions, and behavior. *Social psychology quarterly*, *55*(2),178-204.
[100] Fornell C, & Larcker DF. (1981). Evaluating structural model with unobserved variables and measurement errors. Journal of Marketing Research, *18*(1): 39-50.
[101] Huang, Z., & Benyoucef, M. (2013). From e-commerce to social commerce: A close look at design features. *Electronic Commerce Research and Applications*, *12*(4), 246-259.

# AUTHOR PROFILES

**Adel Bahaddad** received his B.S. degree in Computer Science from King Abdulaziz University, Saudi Arabia in 2002, and received M.S. and PhD degrees in information technology from from School of Information & Communication Technology, Griffith University, Brisbane, Australia in 2011 and 2017 respectively. He is currently work as an assistant professor, Information System department in the Faculty of Computing and Information Technology at King Abdulaziz University. His research interests are mainly focus on e-commerce, e-government and entries architecture.

**Rayed AlGhamdi** received his B.S. degree in Computer Science from Jeddah Teachers' College in 2003, Saudi Arabia. He received M.S. and PhD in Information & Communication Technology from School of Information & Communication Technology, Griffith University, Brisbane, Australia in 2008 and 2014 respectively. He is currently an assistant professor with the Faculty of Computing and Information Technology at King Abdulaziz University (KAU). He serves as a Consultant for Teaching and Learning Development at KAU. His research interests involve E-Systems, technology adoption and IT curriculum.

**Seyed Mohamed Buhari** is an associate professor at Information Technology department, faculty of computing and IT, King Abdulaziz University, Jeddah, Saudi Arabia. He received his PhD qualification in computer networks from Multimedia University, Malaysia, in 2003. Since then, he has gained a number of professional certifications in computer networking from Cisco. His research interests involve network performance and grid computing.

**Madini O. Alassafi** received his B.S. degree in Computer Science from King Abdulaziz University, Saudi Arabia in 2006, and received M.S. degree in Computer Science from California Lutheran University, United State of America in 2013. He received the PhD degree in Security Cloud Computing in April 2018 from University of Southampton, Southampton, United Kingdom. He is currently work as an assistant professor, Information Technology department in the Faculty of Computing and Information Technology at King Abdulaziz University. His research interests are mainly focus on Cloud Computing and Security, Internet of Things (IoT) Security issues. Cloud Security Adoption, Risks, Cloud Migration Project Management and Cloud of Things Security Threats.

**Ahmad A Alzahrani** received a PhD in computer science from La Trobe University, Australia in 2014, he is currently an assistant professor in the faculty of Computing and Information Technology at King Abdulaziz University. His research interests include pervasive computing and Human-Computer Interaction.